# Rapid and Scalable Fabrication of Low Impedance, 3D Dry Electrodes for Physiological Sensing

Ryan Kaveh[1†]*, Natalie Tetreault[1†], Karthik Gopalan[1], Julian Maravilla[1], Michael Lustig[1], Rikky Muller[1,2], Ana C. Arias[1]

† Equally credited authors

[1] University of California, Berkeley, Berkeley, CA 94708 USA.
[2] Chan-Zuckerberg Biohub, San Francisco, CA 94158 USA.

R. Kaveh

University of California, Berkeley, Berkeley, CA 94708 USA.
E-mail: ryankaveh@berkeley.edu

**Abstract:**

Medical electrophysiological sensors that can study the body and diagnose diseases depend on consistently low impedance electrode-skin interfaces. Clinical-standard wet electrodes use hydrogels and skin abrasion to improve the interface and thus the recorded signal quality. These electrodes are challenging to self-administer and impede in-home care. Wearable dry electrodes are more practical; however, they show higher impedances than wet electrodes and are costly to customize. This work presents a fabrication method for rapidly producing low impedance, anatomically fit dry electrodes that do not require hydrogels. By using electroless copper and gold plating with 3D printing, biocompatible electrodes can be optimized for individuals at a fraction of the cost of existing vacuum deposition-based techniques. Example 3D dry electrodes made with this process are evaluated alongside clinical-standard devices in typical scenarios to compare electrical performance and comfort. The resulting dry electrodes exhibited an average electrode-skin impedance of 66.7 kΩ at 50Hz and DC offset of -20 mV without any hydrogel, which, when area normalized, is within the range achieved by wet electrodes without skin abrasion.

## 1. Introduction

In clinical settings, non-invasive electrophysiological (ExG) recordings such as electrocardiography (ECG), electromyography (EMG), and electroencephalography (EEG) enable expansive brain-machine interfaces [1], arrhythmia sensors [2], and advanced prostheses [3]. These devices rely on a low-impedance interface between conductors and skin, achieved by extensive skin preparation and application of electrodes that are “wet“ with hydrogel. This process typically requires a trained professional but results in recordings with high signal-to-noise ratios (SNR), a requirement for restorative and potentially life-changing treatments. These wet electrode hydrogels ensure consistent and low-impedance electrode-skin contact only until the gel dries out. This drying leads to increased electrode-skin impedance (ESI), which reduces recorded single amplitude and increases susceptibility to power-line interference, ultimately reducing SNR [4]. The skin preparation required for wet electrodes (especially in EEG setups for sleep/epilepsy studies) also often leads to skin-irritation, hair loss [5], and discomfort from residual gel left in hair.

Some have attempted to expand these clinical techniques to everyday users by incorporating general purpose semi-dry and dry electrodes. Semi-dry electrodes use significantly less hydrogel that is either pre-applied or stored in an on-electrode reservoir [6] [7] [8] [9]. These semi-dry electrodes can be more comfortable than wet electrodes and achieve similar ESIs but still require some electrolyte and can suffer from control issues such as over-release (which risks bridging). Fully dry electrodes further increase usability and patient comfort but generally result in higher ESI (>1 MΩs at <250 Hz) relative to wet-electrodes (10-100s kΩs at <250 Hz) [10]. Microneedles, pin electrodes, conductive composites, and conformal electrodes have been used to lower ESI and improve the mechanical stability of dry electrodes [10] [11] [12], but introduce new use or fabrication trade-offs. Microneedles, which pierce the top layer of skin, achieve lower ESIs and enable higher SNR recordings. However, prolonged use of these electrodes may result in lesion formation and introduce risk of infection. Non-contact and conductive composites such as silicone carbon black and silver-glass silicone show the opposite trade-off, achieving greater comfort but suffering from higher ESIs relative to other dry electrodes [10] [11] [13]. Other electrode arrays have used flexible planar structures, machined metals, or metal printed devices to increase electrode compliance, comfort, and possibilities in sensing location [14] [15] [16] [17]. Printed flexible electrode arrays enable high density electrode placement, high resolution 2D designs, high-volume fabrication without vacuum, and comfortable electrode compliance along a single axis, such as around arms [17] [18]. Recent work has focused on making chemically inert arrays out of laser-sintered gold. These 2D arrays are potentially more robust than previous demonstrations of silver and graphene based flexible arrays but still generally require hydrogels to record through hairy surfaces [19] [20]. 3D printing and micromachining of insulating and conductive materials have been used in combination with vacuum deposition processes such as sputtering or evaporation [21] [22] to achieve low impedance scalp electrodes. Ultimately, existing techniques emphasize that no existing single dry-electrode design can target all biopotential signals. Furthermore, existing fabrication techniques for comfortable, anatomically customized electrodes are neither low-cost nor scalable or are limited to specific electrode shapes.

Recent commercial, dry electrode ExG [23] [24] [25] systems employ user-generic electrodes and require hairless recording locations on the wrist, forehead or limbs to improve SNR reliability for very specific target applications across large populations. An exploratory medical device that is accepted by a patient for continuous electrophysiological recordings must be anatomically customized for a sensing location that does not interfere with daily activities. It is also preferable to record through hair without the need for shaving, contacting gels or skin preparation without compromising SNR. In addition to low (ESI) in frequency bands of interest (<500 Hz for ExG), near zero electrode DC offset (EDO) is essential for signal quality. An illustration of such ideal electrodes that do not require skin preparation, are reusable, and are fabricated with scalable methods is given in Figure 1. The fabrication process presented in this work combines additive manufacturing with electroless copper [26] and gold plating to enable the rapid prototyping of any anatomically customized, low ESI, reusable dry electrode. By using 3D printing and electroless plating techniques, this process is significantly quicker, more scalable, and cheaper than existing metal sintering and vacuum deposition-based manufacturing techniques for anatomically fit electrodes. Furthermore, the resulting metallized surfaces are more durable than state-of-the-art printed arrays and allow traditional soldering, which is needed for integration with conventional data processing electronics. Lastly, due to their high effective surface areas that can conform to the recording sites, dry electrodes achieve ESIs on par with clinical wet electrodes without the need of a trained technician or skin preparation.

To establish the range of devices that can be fabricated with this process, two dry sensor topographies were designed as examples for use in typical ExG studies. Pin electrodes with bulbed tips were fabricated for increased patient comfort (Figure 1a) during EEG recordings through hair. Versatile circular electrodes were integrated with flexible 3D printed wristband structures for use with different arm and wrist sizes for ECG and EMG measurements (Figure 1b). When compared to clinically standard, wet electrodes of the same area, they express comparable electrode-skin impedances. The results demonstrate a scalable electrode fabrication process that can rapidly produce anatomically optimized, low impedance, 3D dry electrodes for repeatable physiological recordings over long periods of time (Figure 1c). To the authors knowledge, this is the first work to demonstrate a 3D printing and electroless plating based fabrication process for conformal, dry biomedical electrodes.

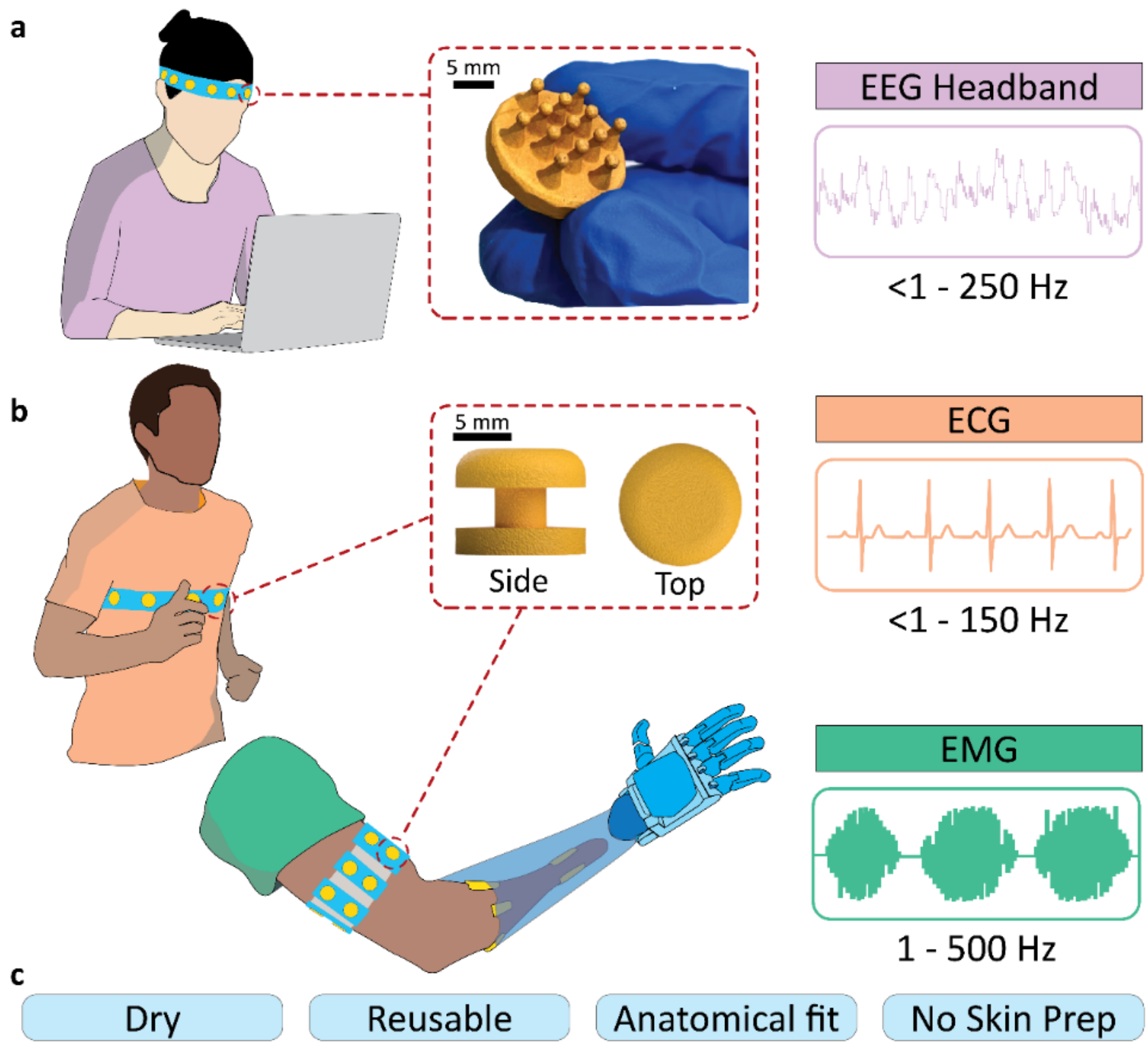

**Figure 1.** Additive manufacturing and electroless plating processes can rapidly fabricate dry electrodes customized for different wearables and partial limbs. This single process can produce designs ranging from dense (a) pin electrodes for EEG recordings through hair to (b) curved/conformal electrodes myoelectric prosthetics at scale. (c) The resulting gold-finished dry electrodes can also be reused over the course of months without any skin preparation.

## 2. Results

### 2.1. Fabrication Process

The electrode structures were printed using a stereolithography 3D printer (Formlabs Form 2 Printer) with a standard, clear methacrylate photopolymer (Figure 2a). Stereolithographic (SLA) printers create 3D structures by precisely laser-curing photosensitive polymer resins in a layer-by-layer fashion. This method of 3D printing results in much finer resolutions and wider choices of materials than fused deposition modeling (FDM) printing, which comprises melted plastic filaments. After printing, the samples were post-processed with a 20-minute IPA bath to rinse uncured resin, followed by an hour-long UV curing process to fully cure the surface.

The structures were subsequently sandblasted with 100 grit white fused aluminum oxide blasting media (Industrial Supply, Twin Falls, ID) to increase their surface area [25] (Figure 2b). Sandblasting promotes better film adhesion and lower skin-electrode impedance. The samples were sonicated in a bath of DI water with Alconox cleaning solution for approximately 10 minutes before rinsing with DI water. The surface energy of the printed structures is modified in a bath of 1% benzalkonium chloride (Sigma Aldrich 12060-100G) surfactant solution for 10 minutes. These steps ensure the plating surface is clean and promote adhesion of the catalyst.

The metal plating process is a result of subsequent plating and cleaning steps. Prior to each plating step, the samples were rinsed and dried thoroughly. First, the 3D printed electrodes were submerged in a beaker of a palladium-tin catalyst for 10 minutes, followed by the copper plating solution for a minimum of 4 hours, which provides a thick base layer of metal for gold plating as shown in Figure 2c. After copper plating, electrodes are soaked in the surfactant and catalyst solutions and then moved to a gold plating solution for approximately 15 minutes (Sigma Aldrich 901670-250ML) (Figure 2d). The second gold plating step is achieved by placing the samples in the same surfactant, catalyst, and gold plating solutions to form the top layer of the dry electrodes (Figure 2e). After all plating steps, tinned copper wires were directly soldered to the electrode surface to facilitate electrode integration with recording systems. This fabrication process ensures two 0.25 μm thick layers of gold and at least 0.5 μm of copper [26] [27]. The thickness of these two layers ensures that the plated surface acts effectively as a short circuit and that any electrode impedance would be dominated by the interface and skin itself. Detailed preparation instructions and processing for catalyst and copper plating solutions can be found below in the methods section.

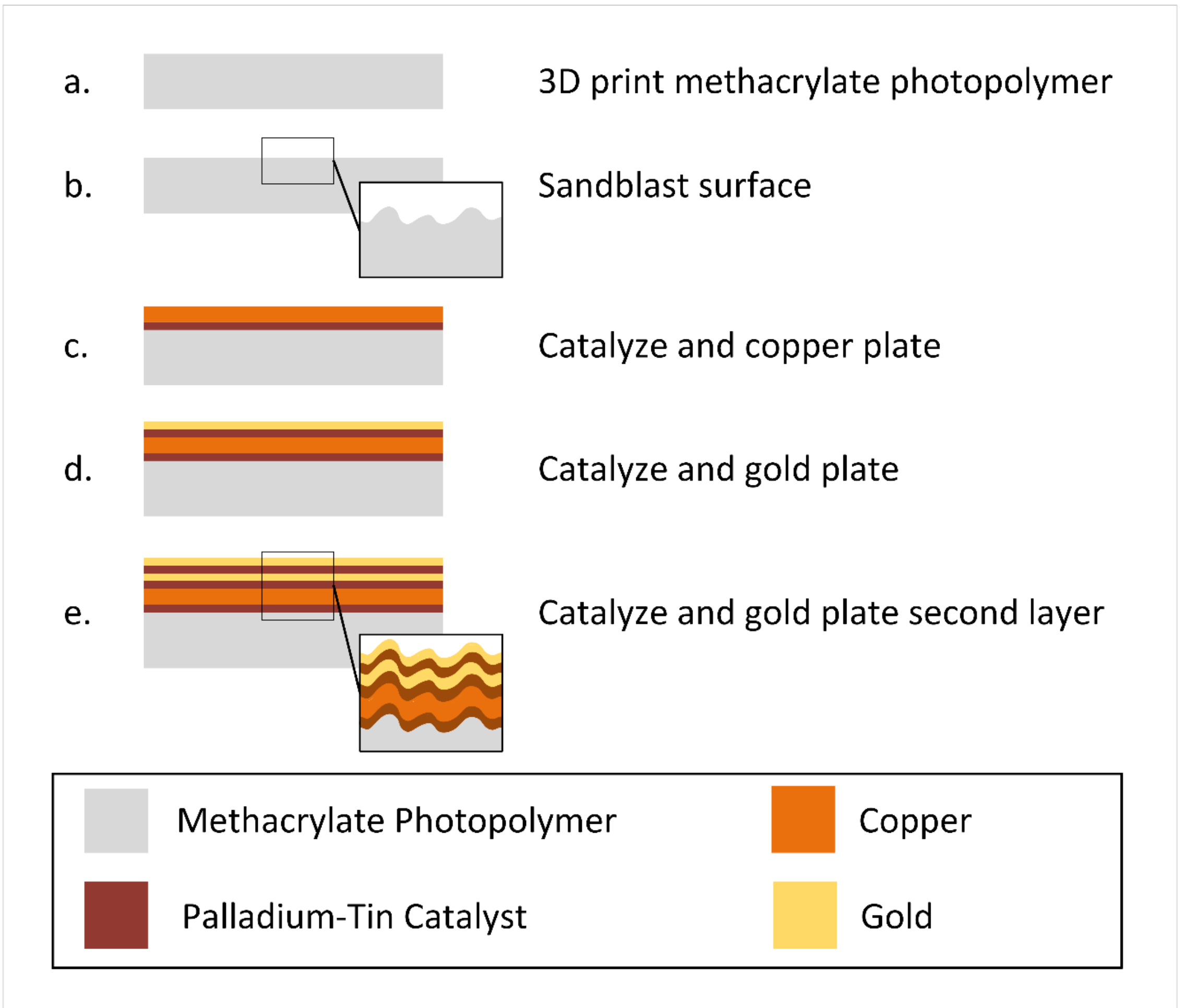

**Figure 2:** Electrode Fabrication Process. (a) Electrode design is 3D printed with an SLA printer. (b) Samples are sandblasted and then cleaned. (c) Electrodes are metalized with copper via exposure to surfactant, catalyst, and copper plating solutions in sequence. (d) The first layer of gold is formed by submersion in surfactant, catalyst, and gold plating solution. (e) The previous step is repeated once more to improve the device longevity and performance.

## 2.2. Material Characterization

### *2.2.1. Sheet Resistance*

Sheet resistance was characterized with 4-point impedance measurements immediately after plating and once a week for nine weeks. Samples of single metal plating layers were compared against double layers, using copper as the control film. As prepared, copper plated control samples, single layer gold samples, and double layer gold samples exhibited an average sheet resistance and standard deviation of 19.2 ± 1.7 mΩ/□, 17.6 ± 1.1 mΩ/□, and 16.5 ± 0.7 mΩ/□, respectively. The second plated layer of gold was shown to lower the sample's average sheet resistance as well as its standard deviation (Figure 3a) and was adopted for the longevity study. It was observed that the deposition of a single layer of gold was not adequate to prevent copper oxidation after a week. After nine weeks, an oxide layer was observed in the copper samples. In addition, a 20% increase in sheet resistance was measured. This contrasts with the 2-layer gold samples which showed no meaningful change in sheet resistance or oxide formation. (Figure 3b). The sheet resistance of both the copper and gold are displayed in reference to their respective sheet resistances on day 0.

### *2.2.2. Film Integrity/Acid Dip Tests*

Kapton tape was applied around entire electrode surfaces and then removed. No visible gold or copper was removed with the tape. Samples also underwent nitric acid baths (Figure 3c). Nitric acid, a typical copper etchant, will readily dissolve copper but is not suitable for etching gold [28]**.** No noticeable differences were observed after dipping the samples with two gold layers. Control samples made of copper were quickly and completely etched down to the bare substrate.

A common issue with microelectronic fabrication is the development of micro or nano cracks from rough handling. The electrodes were observed under optical microscopy after usage and no microcracks were observed at the available scale levels. Additionally, since the electrodes are quite large and rough compared to typical microelectronics components (where microcracks are a common issue), it is unlikely that micro cracks would impact the electrode performance, thus, it was not further studied.

*2.2.2. Surface Energy*

3D printed surfaces require special treatments to accept durable and uniform surface platings. To improve copper adhesion to the 3D printed electrodes, surfaces were sand-blasted and treated with a surfactant. These two steps lead to an increase in surface energy as shown in Figure 3d. Water droplets on untreated surfaces exhibit contact angles of 135$^{o}$ while treated surfaces lead to a significantly smaller contact angle of 64$^{o}$.

*2.2.3. Surface Roughness*

To assess the surface roughness, light microscopy photographs were taken and stylus profilometry was performed on flat, slide-like samples of the same 3D printed photopolymer (Figure 3e & Table 1). The samples were printed, masked with tape, and subjected to the same plating process described above to form selectively patterned films for comparison. A Dektak stylus profilometer was used to collect measurements on these samples at different points in the process and the surface roughness average, $R_a$, was calculated. Due to each plating layer being ≤ 1um thick, the surface roughness was virtually unchanged after the initial sandblasting. Although it may appear that the surface roughness is unchanged before and after sandblasting, the sandblasting creates a more heterogenous surface in all directions as opposed from the highly anisotropic structure of the cured layers of resin. From a user's touch, and from visual inspection under optical microscopy (Figure 3e), change in surface roughness is undetectable after each plating step.

**Table 1.** Sample Surface Roughness at Different Points Along the Plating Procedure

| | Bare print | Sandblasted Surface | Cu | 1 layer Au | 2 layer Au |
|---|---|---|---|---|---|
| Avg Roughness, $R_a$ (μm) | 3.34 ± 0.90 | 3.34 ± 0.67 | 4.32 ± 1.10 | 3.33 ± 1.19 | 3.50 ± 0.82 |

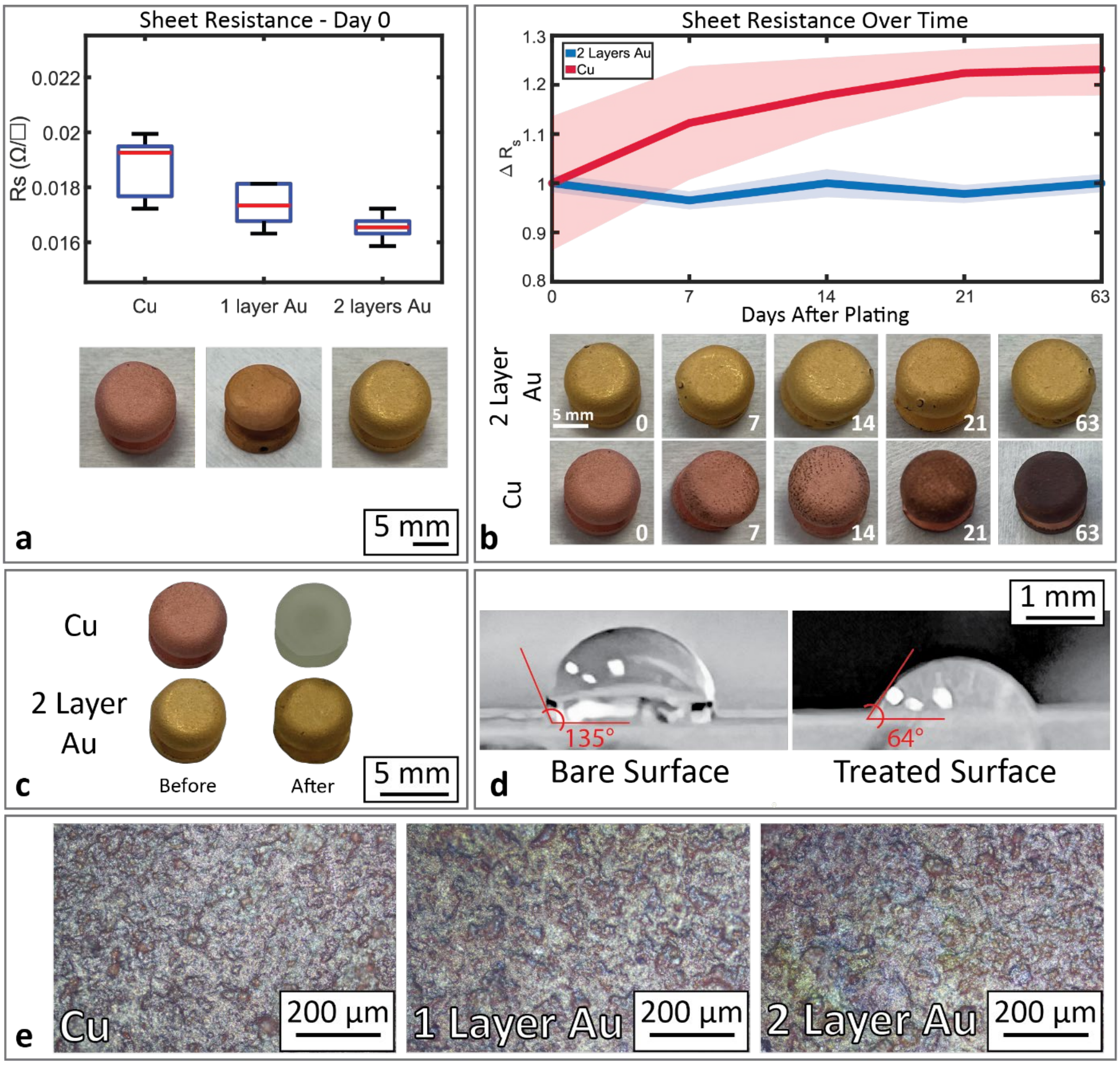

**Figure 3.** Electrical and Physical Characterizations of Au Plated Surfaces. (a) Box plot of absolute sheet resistance immediately after plating. (b) Mean (solid line) sheet resistance and standard deviation (shaded region) over time normalized to day 0. Photos of the copper and copper + two layers of gold samples across the 9-week longevity study. (c) Before and after photographs of acid dip test. The copper sample was etched bare, while the double layered gold sample showed no effect. (d) Contact angle measurements showing the increase in surface energy of 3D printed surface after sand blasting and surfactant treatment. (e) Light microscopy images of plated surfaces showcasing the roughness resulting from sandblasting.

*2.2.4. Electrode Skin Impedance and Electrode DC Offset*

Impedance spectroscopy and EDO measurements with and without electrolyte gel were compared to those of gold cup electrodes. All ESI measurements were performed between two equivalent electrodes, one on the back of the head through the hair, and the other on the ipsilateral mastoid. The presented spectra in Figure 4 are half of the measurement and represent the impedance of one electrode. To emulate real-world scenarios for the dry electrodes, no skin preparation was performed before each trial and measurement sessions were repeated over the course of several weeks. Gold cup electrode measurements used typical skin cleaning prior to each trial. No skin abrasion was performed in order to limit subject discomfort. After donning the electrodes and waiting five minutes for the interface to settle, impedance measurements were performed with an LCR meter (E4980 A, Keysight). Results were fit to an equivalent circuit model using a constant phase element (CPE) (spectra shown in Figure 4a, circuit models shown in Figure 4b and c).

The dry, gold-plated, pin electrodes exhibited an average impedance of 66.7 kΩ and phase of -23° at 50 Hz. When a small amount of electrolyte gel was added to the 3D printed electrodes, the 50 Hz impedance dropped slightly to 41 kΩ while the phase increased to -15°. This is consistent with expected electrode behavior [29]. Comparatively, clinical gold-cup electrodes (without skin abrasion) exhibited a 50 Hz impedance of 31 kΩ and phase of -24°.

Since the dry pin electrodes and wet gold cup electrodes have different contact areas, an area normalized impedance (specific impedance) [7] should be used as a heuristic to compare their performance. The dry electrode had a contact surface area of 50 $mm^2$ and thus a specific impedance an electrode resistivity of 33 $k\Omega cm^2$. The wet electrodes had a total surface area including electrolyte gel of roughly 70 $mm^2$ and thus exhibits a specific impedance of 21.7 $k\Omega cm^2$. Even without skin abrasion, the wet electrode still has a lower resistivity overall, but the dry electrode performance is significantly more area efficient than existing wearable dry electrodes [13]. Furthermore, the dry electrode's gold finish was stable over time. No noticeable surface degradation was observed despite daily reuse and cleaning over the course of two months nor did the measured impedance values increase throughout the observation period (Table 2).

Electrode DC offset (EDO) measurements were taken between each dry sensing electrode and an equivalently sized dry reference electrode with the WANDmini [14] wireless recording device (see Methods section) with an input range up to 400 mV and an input impedance of 10 MΩ. No skin cleaning or abrasion was performed. Initially, the mean EDO and standard deviation was -40 mV ± 15 mV (n=10). After approximately 20 min, the EDO settled to a -20 mV ± 10 mV (n=10). The minimum and maximum EDO values after settling are −32 mV and 23 mV, respectively. It is important to note that the measured EDO is not equivalent to the electrode's open-circuit potential due to the recording front end's finite input impedance.

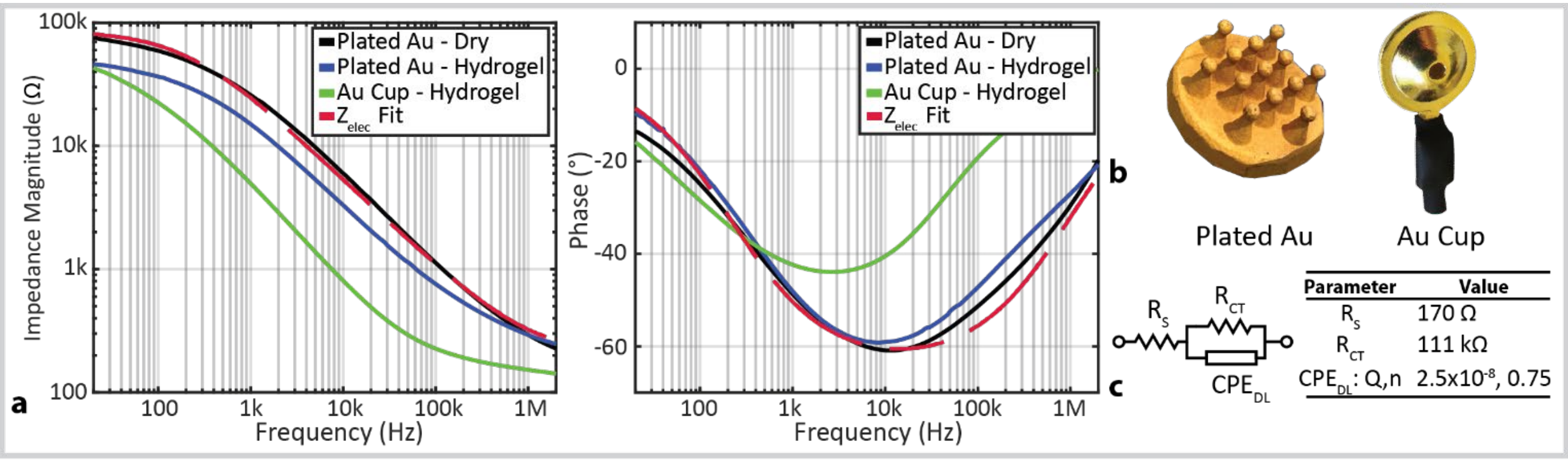

**Figure 4.** Electrode-Skin Interface (ESI) Characterization of Au Plated Pin Electrodes vs Clinical Au Cup Electrodes on across two months. (a) Average (n = 15) magnitude and phase of gold-plated pin electrodes on scalp with and without electrolyte gel alongside gold cup electrodes. An electrode model with a constant phase element (CPE) model, $Z_{elec}$ , was fitted to the dry gold-plated pin electrodes. (b) 3D-printed, gold-plated, dry scalp electrode and clinical gold cup electrode. (c) CPE model with fitted parameters for spread resistance $R_S$, charge transfer resistance $R_{CT}$, and CPE double layer, $CPE_{DL}$

**Table 2.** Au Plated Pin Electrodes Electrode-Skin Interface Impedance over 60 days. Every 30 days the average (n = 5) ESI magnitude and phase were measured. The recording site was neither cleaned nor abraded before measurements. Day to day variation consistent with **[13] [30]** was observed but was likely due to skin condition given visual consistency of electrode coatings.

| Time (Days) | 0 | 30 | 60 |
|---|---|---|---|
| 50 Hz Impedance Magnitude | 86.3 kΩ | 50.1 kΩ | 63.7 kΩ |
| 50 Hz Impedance Phase | -28° | -18° | -23° |

### 2.3. ExG Measurements

ECG, EMG, and EEG signals were recorded using the WANDmini system (see Methods for more details). Three 21-25 year old subjects (one male, two female) were recruited to perform proof-of concept ExG measurements. To mimic realistic day-to-day scenarios, no skin preparation was performed prior to dry electrode donning in all experiments. When wet electrodes were used for comparison measurements, a trained technician performed skin cleaning, hydrogel application, and electrode placement. It is important to note that pressure behind electrodes can result in lower ESI and improve recorded signal quality, but it is unrealistic to expect users to apply great pressure on wearables used on a daily basis. Thus, subjects were tasked with donning the wearable dry electrodes themselves and adjusting the bands until the electrodes were comfortably in contact with their skin. The proof-of-concept study was approved by UC Berkeley's Institutional Review Board (CPHS protocol ID: 2018-09-11395).

*2.3.1. Electrocardiography (ECG)*

Cross body ECG measurements were performed using 3-electrode armbands with embedded curved electrodes on each arm. Differential measurements were taken across both arm bands to record a cross-body ECG (Figure 5a). One of the dry electrodes (Figure 5b) was selected as a ground to reduce interference in the recordings. A clear ECG rhythm (PQRST complex) was recorded and is labeled in Figure 5a.

*2.3.2. Electromyography (EMG)*

Bicep EMG was recorded on a single subject wearing two bands on a single arm, one across the bicep and another by the elbow. The elbow electrodes were used as both a system ground and as references for the bicep electrodes. Sensing electrodes were placed along the bicep. Users were cued to flex their biceps every 5 seconds, resulting in clear increase in broad spectrum (0-200 Hz) electrical activity consistent with large scale muscle activity both in frequency and time domain (Figure 5c, d). The electrode arrangement is shown in Figure 5e.

*2.3.3. Electroencephalography (EEG)*

A single user's alpha attenuation response was measured with both wet and dry electrode-based recording setups. Alpha rhythms are a spontaneous neural signal centered between 8-12 Hz that reflect a person's state of attention and can be modulated by opening and closing their eyes. Originating from the occipital lobe, alpha waves are large amplitude signals that are commonly used to benchmark EEG systems. To make a comparison between EEG recorded with dry and wet electrodes, back-to-back alpha attenuation measurements were performed with the same subject, recording sites, and instrumentation. First, the user would wear a 3D printed headband with pin-style dry electrodes. The headband was oriented to roughly correspond to T3 and T4 recording sides (according to the traditional 10-20 map). Afterwards, the dry electrodes were replaced with clinically standard gold cup electrodes (Figure 4b) and the experiment was repeated. To reduce interference and make the comparison as consistent as possible between the two measurements, a single wet ground electrode was placed on the subject's left mastoid and maintained for both wet and dry electrode measurements. In both setups, the subject was tasked to close/open their eyes every 30 seconds to modulate alpha band activity. When the subject closed their eyes, alpha band power increased while ocular artifacts decreased (Figure 5f). The results showed comparable performance between the wet and dry electrodes. Both wet and dry electrode setups recorded alpha modulation ratios of 4.89 and 4.91, respectively (Figure 5g).

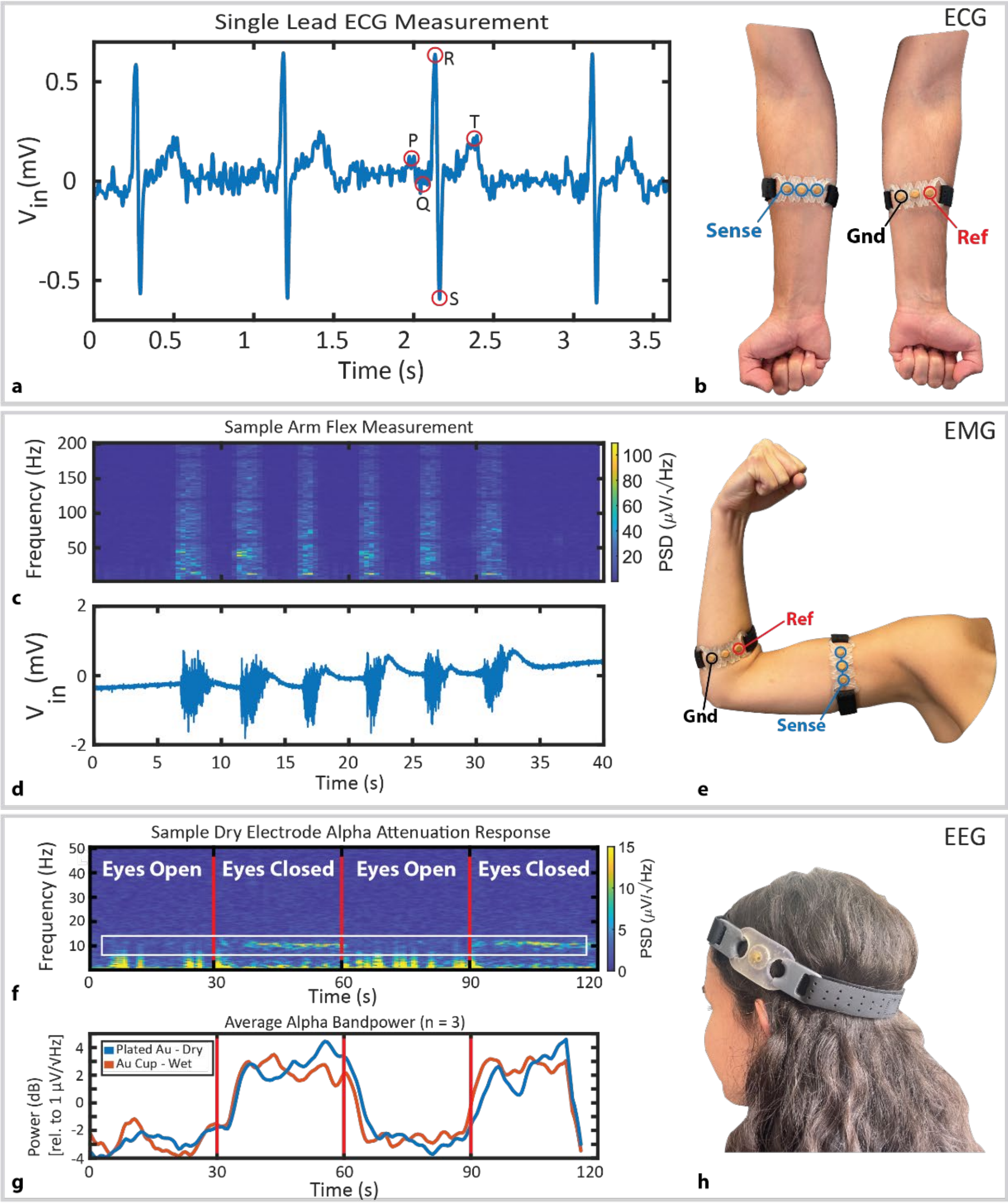

**Figure 5.** Sample physiological measurements taken with gold-plated, dry electrodes. (a) Time domain measurement (n=1) of single lead, cross-body ECG measurement with labeled PQRST complexes. (b) ECG measurement set up with 3D-printed bands with dry electrodes on each arm. (c) Spectrogram and (d) time-domain of EMG measurements (n=1) taken with (e) two electrode bands placed along a single subject's arm. The subject was cued to flex their bicep every 5 seconds. Characteristic broad spectrum muscle activity (~0 - 200 Hz) was observed when the user was cued to flex. (f) Sample Time-frequency spectrogram of a single dry electrode alpha attenuation response. (g) Averaged (n=3) alpha (8-12 Hz) band power of alpha modulation recorded across a subject’s scalp for both dry pin and wet gold cup electrode measurements. Alpha power increased by a factor of 5x in the eyes-closed state for both dry and wet measurements. (h) Alpha attenuation response measurement setup. Two dry electrodes were placed on either side of the head.

*2.3.4. Dry Electrode Comfort*

Subjects donned the curved, dry electrode armband (Figure 5e), dry pin-electrode headband (Figure 5h), and clinically standard, wet, gold cup electrodes (Figure 4b) for three 2 hour sessions. After the third session, subjects asked to rate the comfort of each electrode form factor on a scale of 1-5 (1: unbearably uncomfortable and 5: extremely comfortable). User scores are in Table 3. Wet electrodes had an average rating of 2.3 with subjects citing skin abrasion and residual gel as defining uncomfortable factors. Both dry electrodes were rated 4 on average. The dry electrodes had no skin preparation and left no lingering skin irritation, soreness, or redness. Subjects commented that the curved dry electrodes felt like wearing jewelry and could be worn long term. Wet electrode sites exhibited minor irritation and lingering redness.

**Table 3.** User-specific scoring of the dry electrode headband, dry electrode arm band, and wet electrodes. Dry electrodes were rated significantly more comfortable due to the lack of skin preparation and hydrogel.

| Electrode | User 1 | User 2 | User 3 | Average |
|---|---|---|---|---|
| Pin Dry Electrode | 4 | 4 | 4 | 4 |
| Curved Dry Electrode | 5 | 4 | 4 | 4.3 |
| Wet Electrode | 3 | 2 | 2 | 2.3 |

## 3. Discussion

This multi-layer electroless plating fabrication technique enables the rapid prototyping of customized electrodes for any application. Unlike techniques such as sputtering, evaporation, or spin-coating, the presented electroless plating process will evenly metalize any electrode shape with a gold finish, enabling optimization for any recording location without vacuum or constraints on surface feature density, overhangs, and assembly. The resulting electrode surfaces are robust enough to be directly soldered to, which allows integration with the WANDmini and other systems that use conventional sensor data processing and transmission electronics. Furthermore, this procedure has the added benefit of being performed at relatively low temperatures with standard laboratory equipment, thus preventing carbon scoring or mid-process thin film thermal expansion that can lead to film flaking and peeling (a common issue with electroplating thin films). This increased ease-of-manufacture and ease-of-use also allows for rapid iteration to find comfortable dry electrode designs for specific users and or recording locations (e.g. inside the ear). Since the electroless gold plating process is self-limiting and plates only 0.25um before ceasing, introducing a two-layer stack method increases the thickness. This limits grain-boundary diffusion [31] and extends electrode lifetime, as the electrodes show robust stability over time even after exposure to human skin and ambient air. Tape and acid dip tests further confirm robust gold film integrity and adhesion to the different layers as well as the substrate.

Direct comparisons between the 3D printed, gold-plated dry electrodes and clinical controls are shown by the impedance spectra and EDO measurements with and without electrolyte gel. The ESI, which is crucial to electrode performance, is inversely proportional to the effective electrode surface area [29]. The ESI is further lowered in this process by creating a rough surface via sandblasting, which increases the surface area without increasing the overall size of electrodes [29]. This rough surface is maintained throughout the plating steps as evidenced by the lack of change in surface roughness over each step. The dry, gold-plated electrodes were specifically designed to record through head hair without skin preparation and achieved similar ESI performance to the scalp electrode controls (where a technician cleaned the recording sites and placed wet electrodes) when accounting for the difference in surface area. This similarity emphasizes the performance and ease-of-use gains that can be realized when using electrodes specifically designed for anatomy of the recording site. Lastly, the dry electrodes also maintain EDOs within the input range of state-of-the-art recording front ends [32] [33].

The ECG, EMG, and EEG experiments comprehensively demonstrate the benefits of this optimized fabrication process in sample use cases. In ECG, clear waveforms were shown, demonstrating potential to detect heart rate, heart rate variability, and cardiac arrhythmias. The presented EMG exhibited expected broadband muscle activity. With EEG, the alpha attenuation response recorded with the dry electrodes exhibited the same alpha modulation as the wet electrode case. All three signal paradigms were recorded without any skin cleaning, abrasion, or hydrogel.

Finally, as these electrodes were repeatedly used and tested over the course of two months, there was no degradation in appearance or ESI, nor any adverse skin reaction on users. Ultimately, the presented rapid, adaptable, and low-complexity fabrication process results in re-usable, long lasting, and anatomically fit dry electrodes that can enable new neural wearables and devices for day-to-day brain-computer interfaces.

# 4. Materials and Experimental Section

## 4.1. Plating Solutions

Both the catalyst and copper plating solutions were made in house. The catalyst solution was prepared between 60-70°C and was stirred for approximately one hour after all components were added. This solution is prepared in full and lasts several weeks before the salts precipitate and is no longer usable.

The two main components for an electroless plating solution are a metallic salt and a reducing agent, in this case, copper(II) sulfate and formaldehyde respectively. At a sufficiently high pH (the solution is adjusted to 12.8 by adding NaOH), formaldehyde reacts with hydroxide ions in solution to reduce copper ions in the salt:

$$Cu^{2+} + 2HCHO + 4OH^{-} \rightarrow Cu + H_2 + 2H_2O + 2HCO_2^{-} \quad (1)$$

In the solution we add EDTA, which acts as a complexing agent (as copper salts are insoluble at pH > 4), and ferrocyanide, which acts to stabilize the solution over time (the solution is stored without the addition of formaldehyde, which is then added to the appropriate quantity being used for plating). Layers are typically built at a rate of about <1μm/h at ambient temperature [27], so leaving the samples for several hours or overnight in a covered plating solution provides sufficient coverage. A lightly bubbling nitrogen line was left in the solution to provide light agitation, promote even coverage by displacing the hydrogen gas product, and to limit oxidization of the copper during the plating process.

Lastly, the gold layer was applied by heating the gold plating solution to about 90°C and submerging the samples for about 15 minutes. This process is self-limiting, as gold layers adhere to the copper only and not upon itself. Solution recipes can be found in Table 4.

**Table 4.** Plating process solution components and purpose. All materials were purchased from Sigma-Aldrich.

| Solution | Components | Purpose |
|---|---|---|
| Catalyst | • 1 L deionized water<br>• 60 mL HCl<br>• 0.25g $PdCl_2$ 12g $SnCl_2$, after $PdCl_2$ completely dissolves | Provides very thin initial palladium layer for copper adhesion |
| Electroless Copper | • 1000 mL deionized water<br>• 18g $CuSO_4 \cdot 5H_20$<br>• 48g EDTA<br>• 57.2mg $K_4Fe(CN)_6 \cdot 3H_20$<br>• 1 mL HCl<br>• NaOH as needed to adjust pH to 12.8<br>• Formaldehyde, when ready for use in 22.5:1 ratio of plating solution:formaldehyde | Plates a thick layer of highly conductive material |
| Electroless Gold | Bright electroless gold plating solution (Sigma-Aldrich Part Number: 901670) | Prevents copper oxidation and improves biocompatibility |

**4.2 Experimental ExG Recording System**

In this work, ExG signals are acquired using a miniature, wireless, artifact-free neuromodulation device (WANDmini) [14], a low-profile, custom neural recording system that streams recorded data over Bluetooth Low Energy (BLE) to a base station connected to a laptop (Figure 6). WANDmini is derived from a previous design for a wireless, artifact-free neuromodulation device (WAND) [34], reduced to a form factor of 2.5 × 2.5 $cm^2$, and embedded with custom firmware (Figure 6). Recording and digitization are performed by a custom neuromodulation IC [32] (NMIC, Cortera Neurotechnologies, Inc.) integrated with 64 digitizing frontends, thereby expandable to recording applications with higher electrode counts. NMIC and WANDmini specifications are listed in Table 5.

The NMIC was selected for its low power and high dynamic range, supporting a 100–400 mV input range with a flat input referred noise voltage spectrum of 70 nV/√ Hz. The analog-to-digital converters (ADCs) have a resolution of 15 bits and sample at 1 kSps, providing sufficient resolution and bandwidth for EMG, ECG, and EEG signals. The wide linear input range can accommodate the large electrode DC offsets and provides robustness to interference.

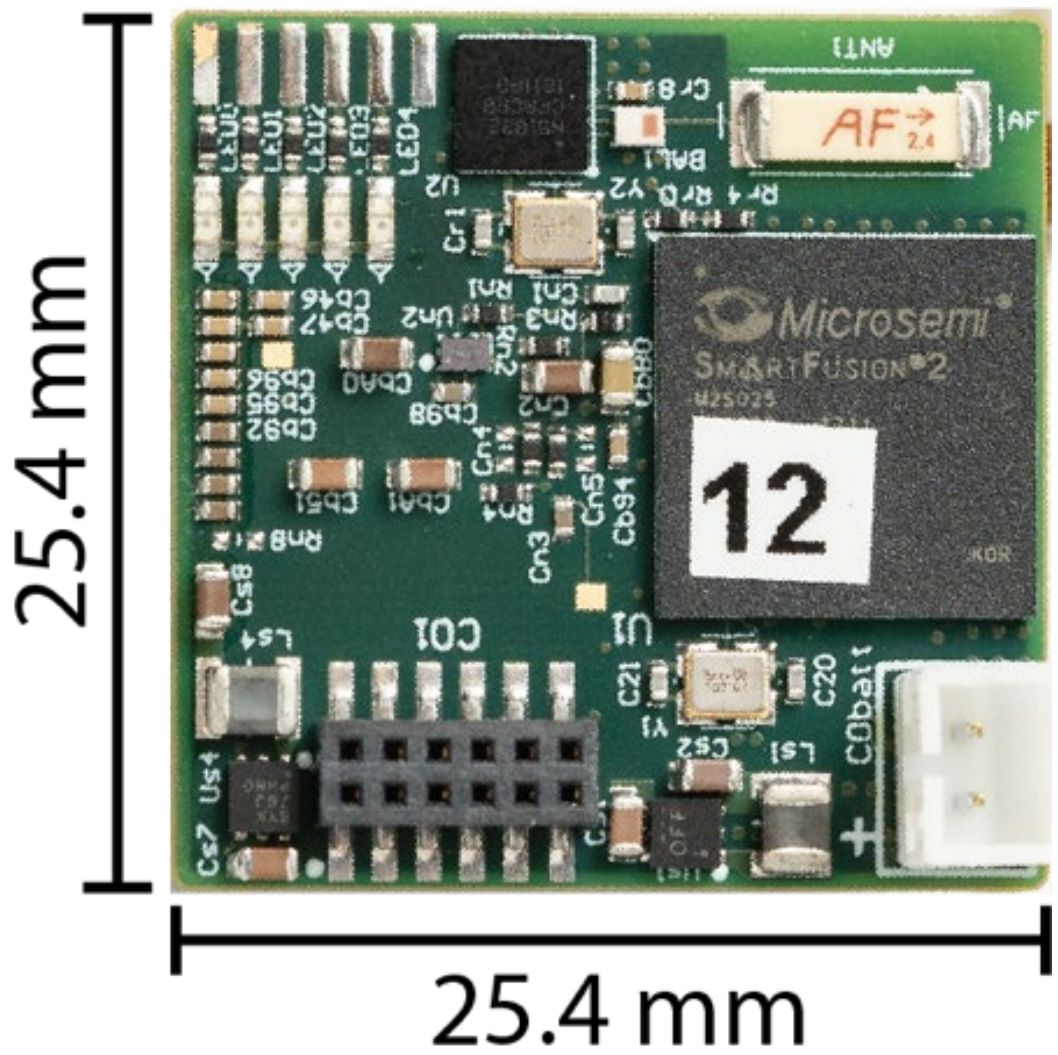

**Figure 6.** WANDmini neural recording module

**Table 5.** Relevant NMIC and WAND mini specifications to for presented ExG measurements.

| NMIC and WANDmini Specifications | |
|---|---|
| Max Recording Channels | 64 |
| Input Range | 100 – 400 mV |
| Input referred noise voltage spectrum | 70 nV/√ Hz |
| Input Impedance | 40 MΩ |
| ADC Resolution | 15 bits |
| ADC Sample Rate | 1 kS/s |
| Wireless Data Rate | 2 Mbps |
| Board Dimensions | 25.4 mm x 25.4 mm |

### 4.3 Analytical Methods

All data presented as mean and standard deviations assume normally distributed data sets with equal variances. For physiological data, minimal pre-processing and normalization was performed. All presented physiological measurements (ECG, EMG, and EEG) were filtered with 60 Hz, 120 Hz, and 180 Hz notch filters using MATLAB. No normalization was performed. ECG in Figure 5a was centered around 0V via mean removal in MATLAB.

**Acknowledgements**

This work was partially supported by NIH U01 EB029427 and the Bakar Fellows Fund. Julian Maravilla acknowledges the NSF for funding under DGE 2146752. The authors would also like to thank the sponsors of the Berkeley Wireless Research Center and Cortera Neurotechnologies for support. Thank you to Sina Faraji Alamouti, Payton Goodrich, Justin Doong, Jasmine Jan, and Emma Wawrzynek for technical discussions.

Received: ((will be filled in by the editorial staff))
Revised: ((will be filled in by the editorial staff))
Published online: ((will be filled in by the editorial staff))

# Supporting Information

**Rapid and Scalable Fabrication of Low Impedance, 3D Dry Electrodes for Physiological Sensing**

Ryan Kaveh*, Natalie Tetreault, Karthik Gopalan, Julian Maravilla, Michael Lustig, Rikky Muller, Ana C. Arias

## Electrode Model Fitting

While many electrode models exist within the literature, CPEs most accurately fit electrode behavior due to their ability to model the imperfect double layer formed at the electrode-skin interface [14]. CPEs are modeled by

$$Z_{CPE} = \frac{1}{(J\omega)^n Q} \quad (1)$$

where $0 < n < 1$. Q is a measure of the magnitude of $Z_{CPE}$ while n fits the bilayer phase offset. The CPE based electrode model's impedance, $Z_{Elec}$, can be described by

$$Z_{Elec} = R_S + \frac{R_{CT}}{1 + (J\omega)^n Q R_{CT}} \quad (2)$$

where $R_s$ is spread resistance and $R_{ct}$ is the charge-transfer resistance. The corresponding circuit model is in Figure 4c.